\documentclass[prl,twocolumn,superscriptaddress,showpacs]{revtex4}

\usepackage{graphicx}
\usepackage{amssymb}
\usepackage{multirow}

\begin{document}

\title{A new type of charged defect in amorphous chalcogenides.}

\author{S. I. Simdyankin}
\email{sis24@cam.ac.uk}
\affiliation{Department of Chemistry, University of Cambridge, 
Lensfield Road, Cambridge CB2 1EW, United Kingdom} 

\author{T. A. Niehaus} 
\affiliation{Fachbereich 6~---~Theoretische Physik, Universit\"at Paderborn, 
             Warburger Stra{\ss}e 100, D-33098, Paderborn, Germany}

\author{G. Natarajan}
\affiliation{Department of Chemistry, University of Cambridge, 
Lensfield Road, Cambridge CB2 1EW, United Kingdom} 

\author{Th. Frauenheim}
\affiliation{Fachbereich 6~---~Theoretische Physik, Universit\"at Paderborn, 
             Warburger Stra{\ss}e 100, D-33098, Paderborn, Germany}

\author{S. R. Elliott}
\affiliation{Department of Chemistry, University of Cambridge, 
Lensfield Road, Cambridge CB2 1EW, United Kingdom} 

\date{\today}

\begin{abstract}
We report on density-functional-based tight-binding (DFTB) simulations of a
series of amorphous arsenic sulfide models. 
In addition to the charged coordination defects previously proposed to exist
in chalcogenide glasses, a novel defect pair, [As$_4$]$^-$-[S$_3$]$^+$,
consisting of a four-fold coordinated arsenic site in a seesaw configuration
and a three-fold coordinated sulfur site in a planar trigonal configuration,
was found in several models.  
The valence-alternation pairs [S$_3$]$^+$-S$_1^-$ are converted into
[As$_4$]$^-$-[S$_3$]$^+$ pairs under HOMO-to-LUMO electronic excitation.
This structural transformation is accompanied by a decrease in the size of the
HOMO-LUMO band gap, which suggests that such transformations could contribute
to photo-darkening in these materials.
\end{abstract}


\maketitle


Amorphous chalcogenides, i.e. sulfides, selenides and tellurides, are 
distinguished from other materials by their photosensitivity.
These materials exhibit intriguing photo-induced phenomena that are not
observed in their crystalline counterparts.
Some of these unusual phenomena have many potential
applications~\cite{Nalwa_HAEPMD}.
A detailed microscopic understanding of the origin of these
phenomena~\cite{Kolobov_PIMAS}, however, is still lacking.

In the context of photo-induced structural changes, special significance is
attributed to the presence of oppositely charged coordination-defect
(valence-alternation) pairs \cite{Kastner_1976}.
Normally, such defect pairs contain singly-coordinated chalcogen atoms having
distinct spectroscopic signatures \cite{Kastner_1976}. 
Experimentally, the concentration of these defects is estimated to be rather
small~\cite{Feltz_AIMG}, i.e. 10$^{17}$ cm$^{-3}$, compared with the atomic
density of about $2\times10^{25}$~cm$^{-3}$, in order quantitatively to
account for the observed magnitude of the photo-induced effects.

Here, we investigate the structure of, and simulate photo-structural changes
in, a series of computer-generated models of the archetypal amorphous
chalcogenide, diarsenic trisulfide (a-As$_2$S$_3$).
In addition to the well-known \cite{Kastner_1976} valence-alternation pairs
(VAPs) [C$_3$]$^+$-C$_1^-$ and P$_4^+$-C$_1^-$, where C$_n$/P$_n$ stands for
an $n$-fold coordinated chalcogen/pnictogen (e.g. sulfur/arsenic) atom, some
of the models were found to contain the previously unreported
[P$_4$]$^-$-[C$_3$]$^+$ defect pairs.
These defect pairs are unusual in two ways. First, there is an excess of
{\em{negative}} charge in the vicinity of the normally electropositive pnictogen
atoms and, second, there are no under-coordinated atoms with dangling bonds in
these local configurations.
The latter peculiarity may be the reason why such defect pairs have not yet
been identified experimentally.

All our simulations have been done by using a density-functional-based
tight-binding method \cite{Frauenheim_2002}, unless specified otherwise.
We found that a basis set of s, p and~d Slater-type orbitals for all atoms is
an essential prerequisite for the observation of the defect-related effects
reported here \cite{Simdyankin_Sicily_2004}.
Each of our models contained 60 (24 As and 36 S) atoms.
This size of model was chosen so as to be big enough that the defect pairs
could be accommodated within the volume of the periodic simulation box, and
yet small enough to allow for the creation and analysis of several
statistically independent models in a reasonable time.
The quality of the models, in terms of comparison with experimental
neutron-diffraction data, is similar to that of the larger models reported in
Ref.~\onlinecite{Simdyankin_2004}.

By using NVT (constant number of particles, volume, and temperature)
molecular-dynamics simulations, we first created a 24-atom model of amorphous
arsenic by quenching from the melt.
Three-fold coordination was imposed by iteratively modifying the
nearest-neighbor shell of each atom and bringing the system to thermal
equilibrium at room temperature.
A stoichiometric model (model 0, in the following) of a-As$_2$S$_3$ was then
created by decorating each of the As-As bonds with S atoms, followed by
rescaling the model to the experimental density \cite{Lee_2004} $\rho =
3.186$~g/cm$^3$ and relaxing at room temperature ($T=300$~K).
This model was then melted at $T=3000$~K and ten more models were created by
taking snapshots of the 3000~K molecular-dynamics trajectory at irregular time
intervals as starting configurations, and following the cooling schedule
described in Ref. \onlinecite{Simdyankin_2004}.

\begin{table}
\caption{Defect statistics for the models containing coordination defects.
Each triplet of numbers denotes, respectively, (I) the number of defects in
the as-prepared ground state, (II) HOMO-LUMO-promoted excited state, and (III)
the ground state resulting from the excited state.}
\begin{tabular*}{8cm}{@{\extracolsep{\fill}}r*{6}{c}}
 \hline \hline
\multirow{3}{*}{Model} & \multicolumn{6}{c}{Number of} \\
       & \multicolumn{4}{c}{coordination defects} & 
           \multicolumn{2}{c}{homopolar bonds}\\ 
   & As$_2$& As$_4$& S$_1$ & S$_3$ & As-As & S-S   \\ \hline
 0 & 0,1,0 & 0,0,0 & 0,1,0 & 0,0,0 & 0,0,0 & 0,0,0 \\ 
 1 & 0,1,0 & 1,1,1 & 0,0,0 & 1,0,1 & 1,1,1 & 1,1,1 \\ 
 2 & 0,2,0 & 0,0,1 & 1,0,0 & 1,0,1 & 2,1,2 & 2,2,2 \\ 
 3 & 0,0,0 & 1,1,1 & 1,1,1 & 0,0,0 & 1,1,1 & 0,0,0 \\ 
 5 & 0,0,0 & 0,0,0 & 0,1,0 & 0,0,0 & 1,1,1 & 1,0,1 \\ 
 6 & 0,0,0 & 1,0,1 & 0,1,0 & 1,1,1 & 2,2,2 & 2,2,2 \\ 
10 & 0,1,0 & 0,0,1 & 1,0,0 & 1,1,1 & 1,0,1 & 1,1,1 \\ 
\hline \hline
\end{tabular*}
\label{tab:defects}
\end{table}

The defect statistics for the models containing coordination defects are shown
in Table~\ref{tab:defects}.
The coordination numbers were calculated as the number of nearest neighbors
within a spherical shell of a radius corresponding to the position of the
first minimum in the corresponding partial radial distribution function.
The numbers in each triplet in Table~\ref{tab:defects} show, respectively, the
number of corresponding defects in (I) the as-prepared ground-state optimized
geometry, (II) the excited-state geometry optimized under the constraint of
having one electron in the highest occupied molecular orbital (HOMO) and one
electron in the lowest unoccupied orbital (LUMO), and (III) the ground-state
optimized geometry obtained from the excited-state configuration.
Although electronic excitations where one electron is promoted from HOMO to
LUMO Kohn-Sham states \cite{Drabold_PRL_99} are not especially realistic, we
simulate such excitations in order qualitatively to assess defect stability
with respect to (optically-induced) electronic excitations.
Note that in models 2 and 10, S$_3$-S$_1$ defect pairs converted into
As$_4$-S$_3$ pairs as a result of the electronic excitation.

The statistics of Mulliken charges (in atomic units) were calculated for all
11 models in the ground state before and after electronic excitation
Since all extremal charge values, which significantly deviate from the average
values (0.32 for As, -0.21 for S), correspond to coordination defects, these
data confirm that the coordination defects form charged defect pairs.
The maximum positive charge (0.46) is on a four-fold coordinated As atom that
is part of an intimate valence-alternation pair (IVAP) where a singly
coordinated S atom is covalently bonded to it.
The minimum positive charge (0.07) is on an As atom that forms a homopolar
bond with an As$_4$ center (see Table~\ref{tab:As4S3charges}).
The minimum negative charge (-0.36) is on a singly coordinated S atom, and the
maximum negative charge (-0.03) is on a S atom that forms a homopolar bond
with another S atom with a similar charge value.

It is common \cite{Kastner_1976} to use the following notation for
valence-alternation pairs: As$_4^+$-S$_1^-$ and S$_3^+$-S$_1^-$, where the
superscripts show the polarity of the nominal excess charge on corresponding
defect centers.
Although the charges on As$_4$ and S$_1$ atoms are of correct polarity in an
alloy of normally electropositive As and electronegative S, those on S$_3$ are
not.
Therefore, it is more appropriate to use the notation [S$_3$]$^+$ for
three-fold coordinated S atoms, where the square brackets imply that the
excess charge is distributed between the center and its nearest neighbors.
Table~\ref{tab:As4S3charges} shows that there is also an excess negative
charge in the vicinity of As$_4$ centers when [S$_3$]$^+$ centers are also
present, and these newly-identified defects will be denoted as [As$_4$]$^-$
here.

\begin{table*}
\caption{Mulliken charges, in atomic units, on atoms within the
[As$_4$]$^-$-[S$_3$]$^+$ defect pairs. The index (I) next to the model number
signifies the as-prepared ground-state geometry, and (III) indicates the
ground-state structure obtained by geometry optimization of the excited-state
configuration (II).  The relevant fragment of model 2(III) and clusters
[SAs$_3$H$_6$]$^+$ and [AsS$_4$H$_4$]$^-$ are depicted in
Fig.~\ref{fig:as4_s3}(a)-(c), respectively.}
\begin{tabular*}{16cm}{@{\extracolsep{\fill}}*{7}{r}}
 \hline \hline
\multirow{2}{*}{Model} & \multirow{2}{*}{Center} &
\multicolumn{4}{c}{Neighbors} & \multirow{2}{*}{Total} \\
 & & 1 & 2 & 3 & 4 &  \\ \hline
\multirow{2}{*}{1(I)} & 
(As) 0.313 & (S) -0.194 & (S) -0.281 & (S) -0.223 & (S) -0.313 & -0.698 \\ 
                      & 
(S) -0.110 & (As) 0.363 & (As) 0.379 & (As) 0.362 &            &  0.994 \\
\multirow{2}{*}{2(III)} & 
(As) 0.329 & (As) 0.071 & (S) -0.221 & (S) -0.203 & (S) -0.329 & -0.353 \\ 
                      & 
(S) -0.153 & (As) 0.355 & (As) 0.347 & (As) 0.351 &            &  0.900 \\
\multirow{2}{*}{6(I)} & 
(As) 0.345 & (S) -0.149 & (S) -0.255 & (S) -0.295 & (S) -0.225 & -0.579 \\ 
                      & 
(S) -0.146 & (As) 0.375 & (As) 0.376 & (As) 0.384 &            &  0.989 \\
\multirow{2}{*}{10(III)} & 
(As) 0.326 & (As) 0.101 & (S) -0.242 & (S) -0.308 & (S) -0.231 & -0.354 \\ 
                      & 
(S) -0.158 & (As) 0.363 & (As) 0.295 & (As) 0.101 &            &  0.601 \\
\textrm{[AsS$_4$H$_4$]$^-$} &
(As) 0.280 & (S) -0.359 & (S) -0.320 & (S) -0.498 & (S) -0.600 & -1.497 \\
\textrm{[SAs$_3$H$_6$]$^+$} & 
(S) -0.149 & (As) 0.365 & (As) 0.362 & (As) 0.374 &            &  0.952 \\
\hline \hline
\end{tabular*}
\label{tab:As4S3charges}
\end{table*}

\begin{figure} 
\centerline{\includegraphics[width=6cm]{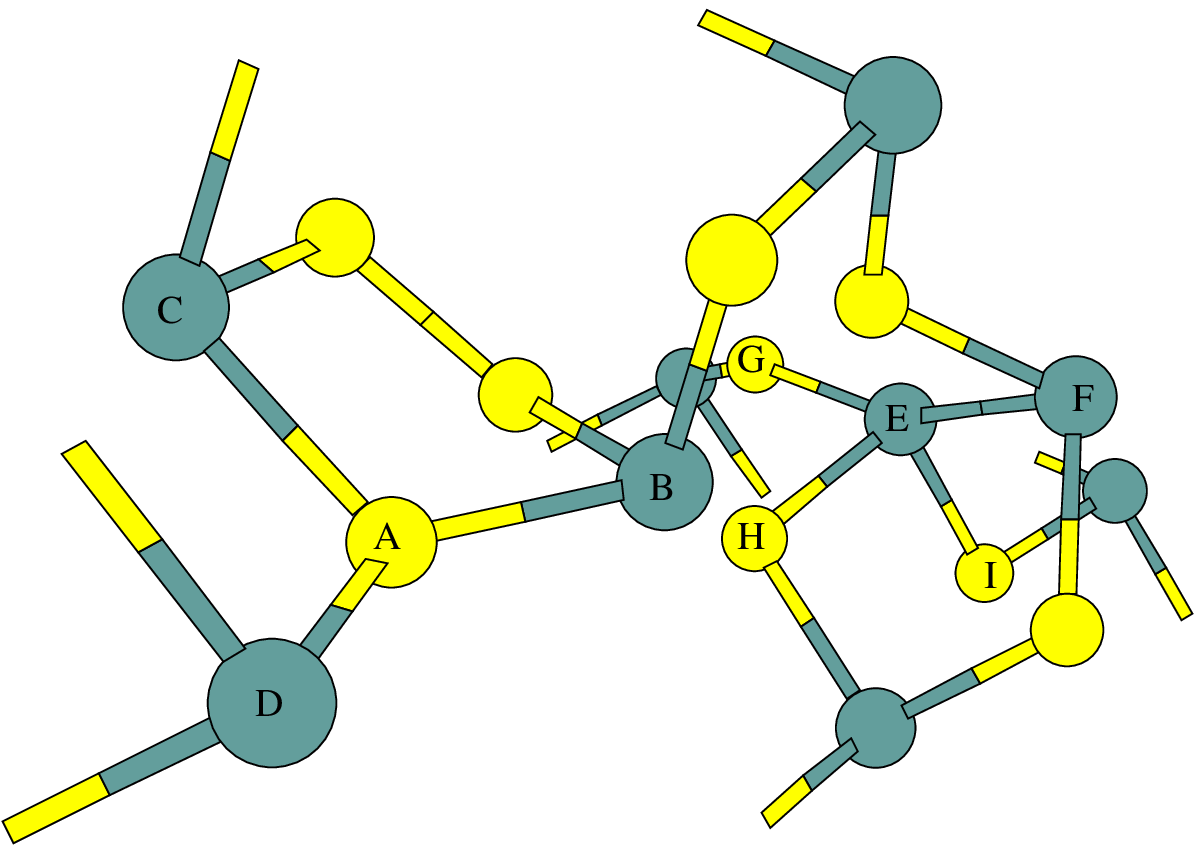}} 
\centerline{(a)}
\centerline{\hfill \includegraphics[width=3.2cm]{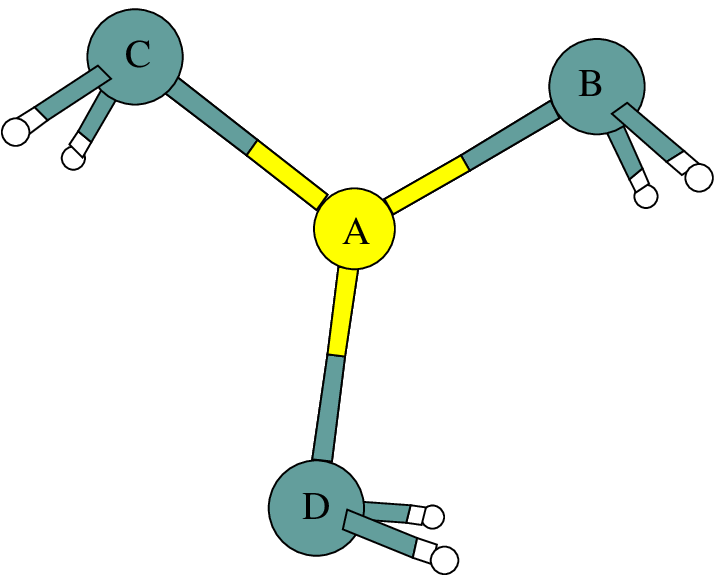} \hfill
                   \includegraphics[width=2.8cm]{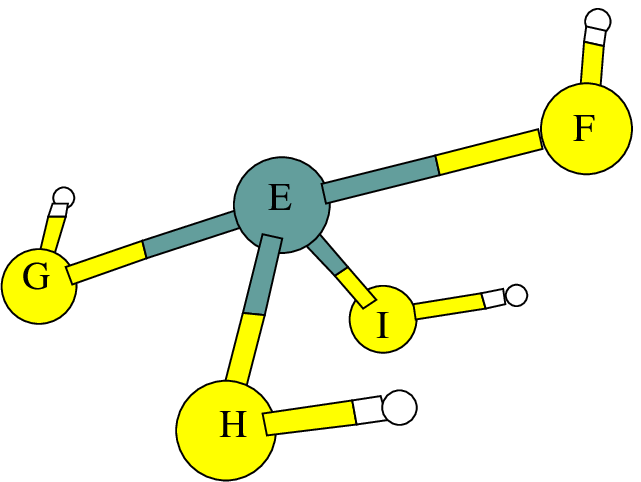} \hfill} 
\centerline{\hfill (b) \hfill (c) \hfill}
\caption{Planar trigonal, [S$_3$]$^+$ (marked by letter `A'), and seesaw,
[As$_4$]$^-$ (marked by letter `E'), configurations in (a) a fragment of model
2(III) (the dangling bonds show where the displayed configuration connects to
the rest (not shown) of the network) and, (b) and (c), charged isolated
clusters (the dangling bonds are terminated with hydrogen atoms).  The shading
of the As atoms is darker than that of the S atoms. Bond lengths are (\AA):
(a) AB=AD=2.42, AC=2.39, EF=2.83, EG=2.54, EH=2.35, EI=2.33; (b) AB=2.34,
AC=AD=2.33; (c) EF=2.55, EG=2.46, EH=2.30, EI=2.32. Color online.}
\label{fig:as4_s3}
\end{figure}

\begin{figure} 
\centerline{\hfill \includegraphics[width=4cm]{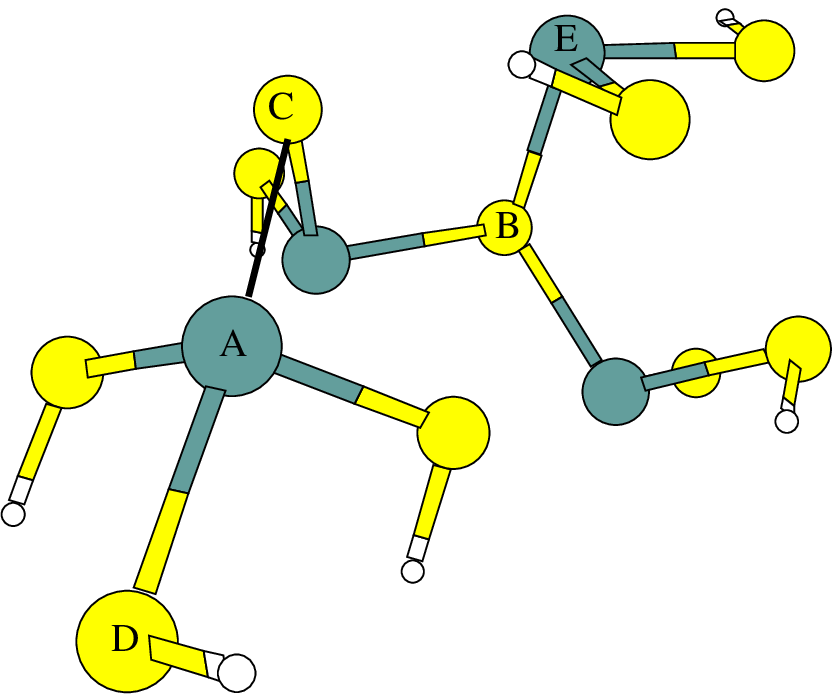} \hfill 
            \includegraphics[width=4cm]{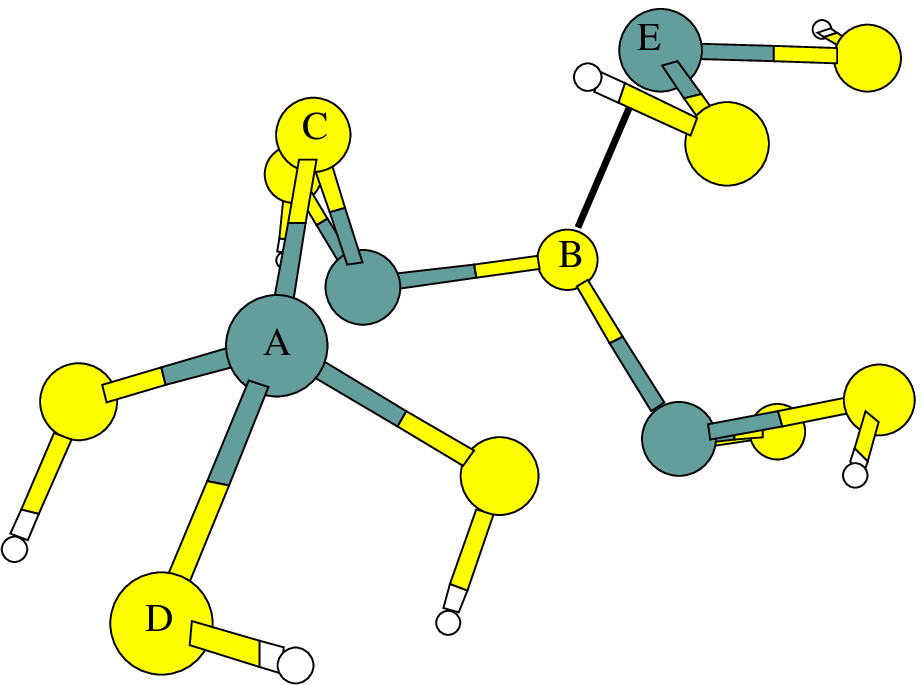} \hfill}
\centerline{\hfill (a) \hfill $\;$ \hfill (c) \hfill}
\centerline{\hfill \includegraphics[width=4cm]{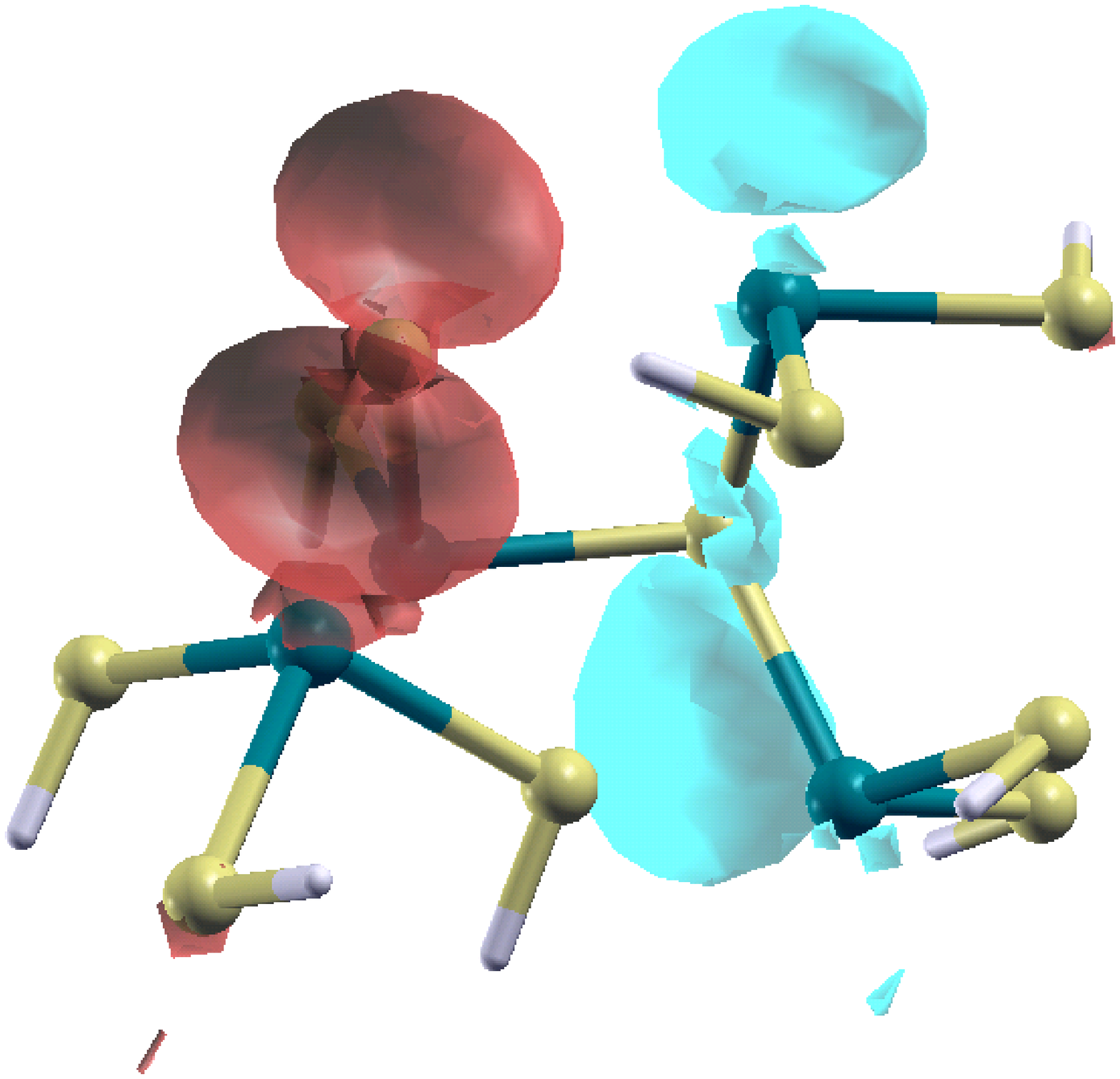} \hfill
            \includegraphics[width=4cm]{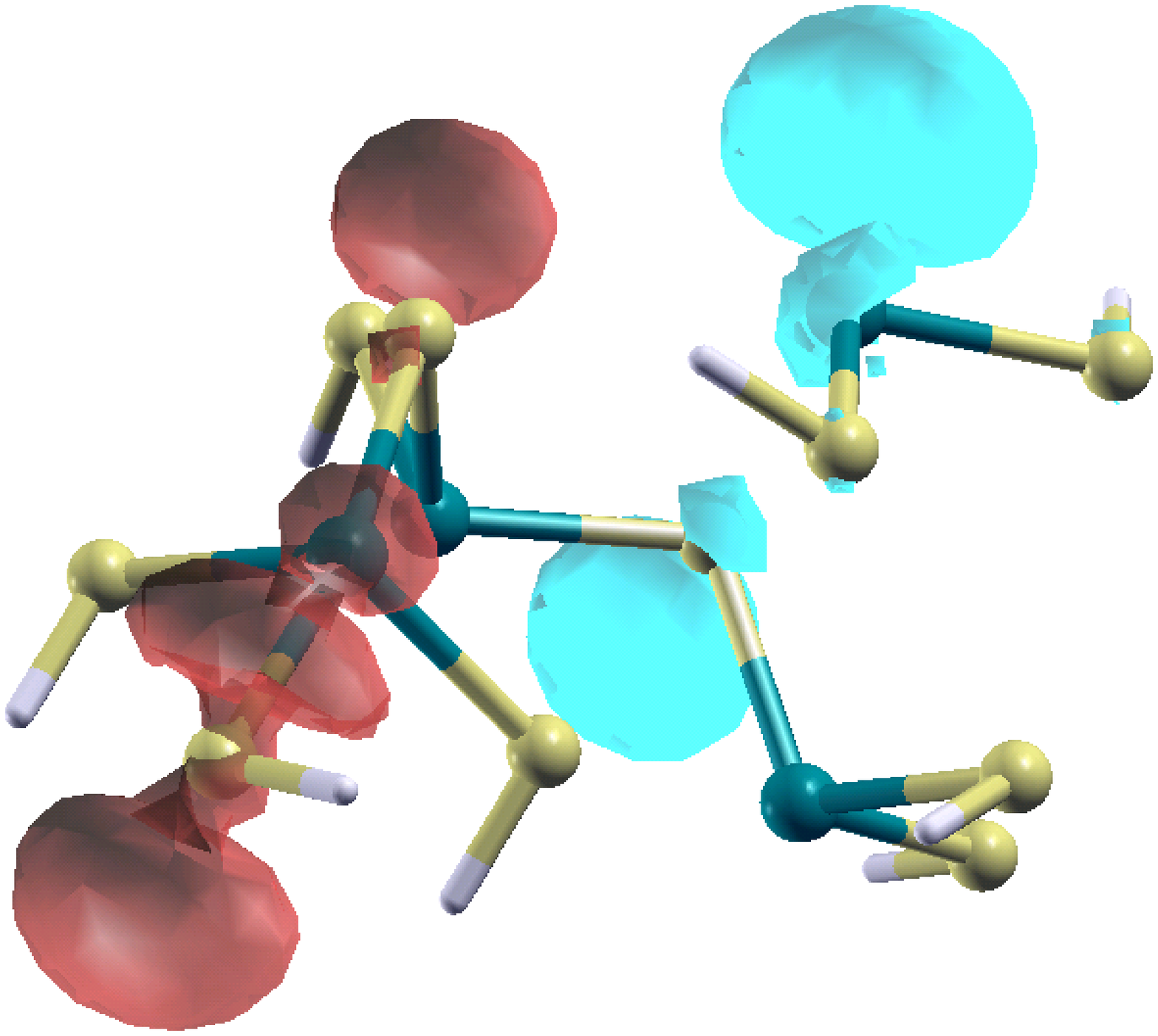} \hfill}
\centerline{\hfill (b) \hfill $\;$ \hfill (d) \hfill}
\caption{An As$_4$S$_{10}$H$_8$ cluster containing both defect centers
[As$_4$]$^-$ (marked by letter `A') and [S$_3$]$^+$ (`B'). The shading of the
atoms is the same as in Fig~\ref{fig:as4_s3}. The black solid lines signify
elongated bonds.  (a) Optimized ground-state geometry.  Bond lengths are
({\AA}): AC=3.00, AD=2.32, and BE=2.4.  (b) Isosurfaces corresponding to the
value of 0.025 of electron density in the HOMO (darker red surface) and LUMO
(lighter cyan surface) states for the structure shown in (a).  (c) Optimized
geometry in the first singlet excited state. Bond lengths are ({\AA}): AC~=
2.43, AD~= 2.44, and BE~= 2.82.  (d) Same as (b), but for the structure shown
in (c).  Color online.}
\label{fig:as4_s3_cluster}
\end{figure}

It is of interest to verify whether the excess charge has a similar effect on
the local structure in both super-cell amorphous and isolated cluster models.
Fig.~\ref{fig:as4_s3}(a) shows a fragment of model 2(III) containing both
[As$_4$]$^-$ and [S$_3$]$^+$ defects.
Fig.~\ref{fig:as4_s3} also shows the optimized geometry in (b)
[SAs$_3$H$_6$]$^+$ and (c) [AsS$_4$H$_4$]$^-$ clusters, where one electron was
removed from and added to the neutral system, respectively.
The seesaw [As$_4$]$^-$ and planar [S$_3$]$^+$ local configurations observed
in models 1(I), 2(III) (see Fig.~\ref{fig:as4_s3}(a)), 6(I) and 10(III) are
indeed very similar to those in the isolated clusters shown in
Fig.~\ref{fig:as4_s3} (b) and (c), respectively.
We note that similar seesaw-like structures are also observed in local
configurations around hypervalent chalcogens, e.g. in sulfur(IV) fluoride and
in certain tellurides \cite{McLaughlin_2000}, and are expected from
stereochemical considerations \cite{North_PASC}.
Also, the five-fold coordinated As defect reported in \cite{Uchino_2000} can
be viewed as an [As$_4$]$^-$ center bonded to an additional As atom (placed on
top of the structure depicted in Fig.~\ref{fig:as4_s3}(c)).
Interestingly, five-membered rings fit conveniently to parts of the
over-coordinated defect centers, which might contribute to the stability of this
type of defect pair.
It is also worth mentioning that there is a distinct pattern in the
distribution of charges in the seesaw configuration, which is most clearly
seen in the isolated [AsS$_4$H$_4$]$^-$ cluster,~--- the more distant sulfur
atoms F and G (see Fig.~\ref{fig:as4_s3}(c)) have the largest magnitude negative charges (see
Table~\ref{tab:As4S3charges}).

Although the DFTB method involves a number of approximations, it has been
shown \cite{Frauenheim_2002} that its accuracy is comparable to that of full
DFT methods.
In order to verify whether the [As$_4$]$^-$-[S$_3$]$^+$ defect pair is
reproducible with more accurate methods, we optimized the geometry of
model~1(I) by SIESTA \cite{Soler_02}, which is a DFT method using an
atomic-orbital basis set and pseudopotentials to eliminate the core electrons.
Apart from insignificant changes in bond lengths and angles, the SIESTA- and
DFTB-optimized models are very similar, including the region with the
seesaw-planar trigonal defect pair.
We also performed high-accuracy all-electron geometry optimization of the
isolated charged clusters shown in Fig.~\ref{fig:as4_s3}(b) and (c) by using
NWChem \cite{NWCHEM1, NWCHEM2}. The calculations were performed at the
all-electron DFT level with a B3LYP hybrid exchange-correlation functional and
a 6-311G** gaussian basis set which includes s, p and d orbitals on the sulfur
and arsenic atoms.  The optimized DFTB geometry was used as a starting
configuration for the all-electron optimizations. After optimization, the
shapes of the molecular clusters were preserved, apart from a slight increase
in the bond lengths, EF=2.80~{\AA} and EG=2.51~\AA, in the ``seesaw''
configuration.

We have found that the hitherto unreported [As$_4$]$^-$-[S$_3$]$^+$ defect
pairs in our models are stable with respect to HOMO-to-LUMO electron
excitations.
We also performed a geometry optimization of an As$_4$S$_{10}$H$_8$ cluster,
containing an [As$_4$]$^-$-[S$_3$]$^+$ pair by construction, both in the
ground state and in the first singlet excited state within the linear-response
approximation to time-dependent density-functional theory, which gives a much
better description of excited states compared with HOMO-to-LUMO electron
excitations \cite{Niehaus_2001} (see Fig.~\ref{fig:as4_s3_cluster}).
Although the bond AC in the ground-state structure depicted in
Fig.~\ref{fig:as4_s3_cluster}(a) is significantly elongated, analysis of the
electron density in the HOMO and LUMO electronic states (see
Fig.~\ref{fig:as4_s3_cluster}(b)) shows that it has a significant bonding
character, implying that this cluster contains a distorted
[As$_4$]$^-$-[S$_3$]$^+$ pair.
As seen in Fig.~\ref{fig:as4_s3_cluster}(c), redistribution of the electron
density in the excited state leads to a symmetrization of the As$_4$ center
and to an elongation of the bond BE in the region of the S$_3$ center, where
the LUMO electronic state is predominantly localized.
Subsequent geometry optimization of the excited-state configuration results in
the same ground-state geometry as in Fig.~\ref{fig:as4_s3_cluster}(a).
We observed that bond breaking/elongation in all our models generally occurs
at the groups of atoms where the LUMO is localized, indicating the expected
antibonding character of LUMO states.

\begin{table}
\caption{HOMO-LUMO band-gap energies for the all-heteropolar model 0 and the
models containing coordination defects. The indices (I), (II), (III)
correspond to the triplets in Table~\ref{tab:defects}.}
\begin{tabular*}{8.5cm}{@{\extracolsep{\fill}}r*{4}{c}}
 \hline \hline
\multirow{2}{*}{Model} & 
     \multicolumn{3}{c}{Band gap (eV)} \\ 
    & (I)   & (II) & (III) \\ \hline
 0  &  1.46 & 0.68 & 1.46  \\
 1  &  1.13 & 0.94 & 1.13  \\
 2  &  1.41 & 0.62 & 1.06  \\
 3  &  1.23 & 0.21 & 1.24  \\
 4  &  1.47 & 1.50 & 1.47  \\
 5  &  1.68 & 0.40 & 1.67  \\
 6  &  1.47 & 1.37 & 1.47  \\
 7  &  1.47 & 1.50 & 1.47  \\
 8  &  1.38 & 1.20 & 1.38  \\
 9  &  1.65 & 1.31 & 1.65  \\
10  &  1.20 & 0.85 & 1.02  \\
\hline \hline						       
\end{tabular*}
\label{tab:gaps}
\end{table}

The HOMO-LUMO band-gap energies for models 0-10 are listed in
Table~\ref{tab:gaps}.
As in Table~\ref{tab:defects}, the three values correspond, respectively, to
the same ground-, excited-, ground-state geometries.
Note that in models 2 and 10, the magnitude of the band gap has decreased as
[S$_3$]$^+$-S$_1^-$ pairs were converted into [As$_4$]$^-$-[S$_3$]$^+$ pairs.
It should be noted that [S$_3$]$^+$ centers are not necessarily conserved
under such conversions~--- while in model~10 the S$_3$ atom is the same in
both ground states (I) and (III), in model~2 it is not.


In summary, we have demonstrated the existence of a new type of charged defect
pair in amorphous arsenic sulfide, namely [As$_4$]$^-$-[S$_3$]$^+$, where the
As center has a ``seesaw'' configuration, and the S center is trigonal
planar.  A plausible scenario for photo-darkening in this material is the
conversion of other types of defects containing under-coordinated atoms into
such defect pairs due to electronic excitation under
illumination. Interestingly, in the two models where such conversion has been
observed, As-As homopolar bonds were formed as  part of the [As$_4$]$^-$
centers. Given that their bond lengths are greater than the average bond
length in the material, such bond formation could also contribute to
photo-expansion of the material.


S.I.S. and G.N. are grateful to the EPSRC for financial support.  We thank the
British Council and DAAD for provision of financial support.  We used the
XMakemol \cite{xmakemol} and XCrySDen \cite{xcrysden} software for
visualization.

\bibliographystyle{apsrev}

\end{document}